\documentclass[prb,twocolumn,showpacs,preprintnumbers,amsmath,amssymb]{revtex4}

\usepackage{graphicx}
\usepackage{dcolumn}
\usepackage{bm}

\begin{document}
\title{Momentum-dependent resonant inelastic X-ray scattering at the Si
$K$ edge of 3C-SiC:\\ A theoretical study on a relation between spectra and
valence band dispersion}

\author{Yunori Nisikawa}
\email{nisikawa@sci.osaka-cu.ac.jp}
\affiliation{Department of Material Science, Osaka City University, 
Sumiyoshi-ku, Osaka 558-8585, Japan}%
\author{Muneharu Ibuki}
\affiliation{3-2-26 Tezukayama, Nara City, Nara 631-0062, Japan}
%
%
\author{Manabu Usuda}
\affiliation{Masago 1-4-8, Ibaraki, Osaka, 567-0851, Japan}%

\date{\today}

\begin{abstract}

We theoretically demonstrate that a resonant inelastic x-ray scattering
 (RIXS) with a sizable momentum transfer can be utilized to study
 valence band dispersion for broad band materials. We take RIXS at the
 Si $K$ edge of 3C-SiC as a typical example.
The RIXS spectra are calculated by systematically changing the
 transferred momentum, an incident photon polarization and an incident photon　energy, on the basis of an ab initio calculation.
We find that the spectra depend heavily on both the transferred
 momentum and the incident photon polarization, and the peaks in the spectra correspond to the energies of the valence bands. 
We conclude that the information on the energy dispersion of valence
 bands can be extracted from the transferred momentum dependence of 
the RIXS spectra. 
These findings lead to further application for RIXS when investigating the band structure of broad band materials.
\end{abstract}


\maketitle

\section{Introduction}
%
%
%
Inelastic x-ray scattering spectroscopy is a promising method used for studying the electronic structure of matter.
Since the scattering intensity in the inelastic part is much smaller than that in the elastic part, it is necessary to use x-ray optics with a high-brilliance synchrotron radiation source.
It is advantageous to use a resonant enhancement by tuning the photon energy close to the absorption edge.
The resonant inelastic x-ray scattering (RIXS) technique has advantages such as bulk sensitivity and element specificity. The technique can be used to investigate electronic structure in various materials.
%

In semiconductors such as Si and Ge, the RIXS process can be interpreted as follows. A core electron is excited to an unoccupied conduction band state of crystal momentum ${\bf k}_e$ by an incident photon with  momentum ${\bf q}_1$, while a valence electron with crystal momentum ${\bf k}_h$  fills the core hole, emitting a photon with momentum ${\bf q}_2$.
In the final state, electron-hole pairs remain and the momentum is conserved as ${\bf q}_1-{\bf q}_2={\bf k}_e-{\bf k}_h$ in the whole process.
Here, ${\bf q}_1-{\bf q}_2$ is the transferred momentum from the photons to the electronic system.
Therefore the observed RIXS spectrum contains information about the conduction and valence bands.

%
When the magnitude of transferred momentum
$\Delta q \equiv |{\bf q}_1-{\bf q}_2|$
 is much smaller than the typical size of the crystal momentum
$k_{\rm elec}$ of electrons in the solid, 
the momentum conservation can be written as ${\bf k}_e={\bf k}_h$.
By tuning the incident photon energy,　 we can indirectly select
${\bf k}_e$ and thereby ${\bf k}_h(={\bf k}_e)$, depending on the
dispersion of the unoccupied conduction bands.
Therefore, from the excitation energy dependence of RIXS spectra,　we can
determine unambiguously valence band dispersion at special points, such
as band edge and high symmetry point ( k-selective RIXS ).
\cite{rf:SiL23, rf:C-dia, rf:SiL23-1, rf:C-gra, rf:GaN, rf:hBN,  rf:cBN,
rf:3C-SiC, rf:SiL3, rf:p-SiC, rf:BP}
The situation mentioned above usually corresponds to RIXS in the soft X-ray
region.
The k-selective RIXS in the soft X-ray region has been performed for a number of broad band materials
\cite{rf:GaP, rf:CaB6, rf:C-gra-1, rf:MgB2, rf:LiBC, rf:GaN2004,rf:GaN2005, rf:BeTe, rf:CdS}
and is widely accepted as a tool for band structure investigation.
\cite{rf:C-dia-th, rf:C-gra-th, rf:hBN-th, rf:th-1, rf:th-2, rf:BP-th, 
 rf:rev-1, rf:rev-2, rf:rev-3, rf:rev-4, rf:rev-5, rf:rev-6}

In the case of $ \Delta q \simeq k_{\rm elec}$,　
${\bf q}_1-{\bf q}_2$ can cover a wide range of the Brillouin
zone and we can  control its magnitude and direction
 by setting a scattering geometry.
Therefore, we can scan ${\bf k}_h$ as ${\bf k}_h={\bf k}_e-({\bf
q}_1-{\bf q}_2)$, where we can select ${\bf k}_e$ by tuning the incident
photon energy.
We can then extract information on the dispersion of valence bands from the transferred momentum dependence of the RIXS spectra without changing the incident photon energy ( k-scannable RIXS )\cite{rf:Math}.
In contrast to the k-selective RIXS experiment for broad band material, the k-scannable RIXS experiments are far from being exhaustive with the exception of experiments performed for 
Si,\cite{rf:Maexp} Ge,\cite{rf:GeCu, rf:Geth} Cu \cite{rf:GeCu} and NiAl \cite{rf:NiAl}.
In k-scannable RIXS there are two important conditions:
(A)$ \Delta q \gtrsim k_{\rm elec}$ 
(B) $\Gamma \ll W_{\rm con}$, where $\Gamma$ is core level width and 
$W_{\rm con}$ is the bandwidth of the lowest conduction band.
Condition (A) ensures the scannability while condition (B) ensures the
selectivity of ${\bf k}_e$. These conditions also make sure the RIXS spectra is dependent on transferred momentum.
( It should be mentioned that the photon energy $E$ is required to obtain
sizable momentum transfer $\Delta q\sim E/\hbar c$ but corresponding
core level widths $\Gamma$ are usually large.)
For example, the k-scannable RIXS spectra at the $K$ edge of Ge are
nearly independent of transferred momentum.\cite{rf:GeCu, rf:Geth}.

\begin{figure}
\includegraphics[width=0.80 \linewidth]{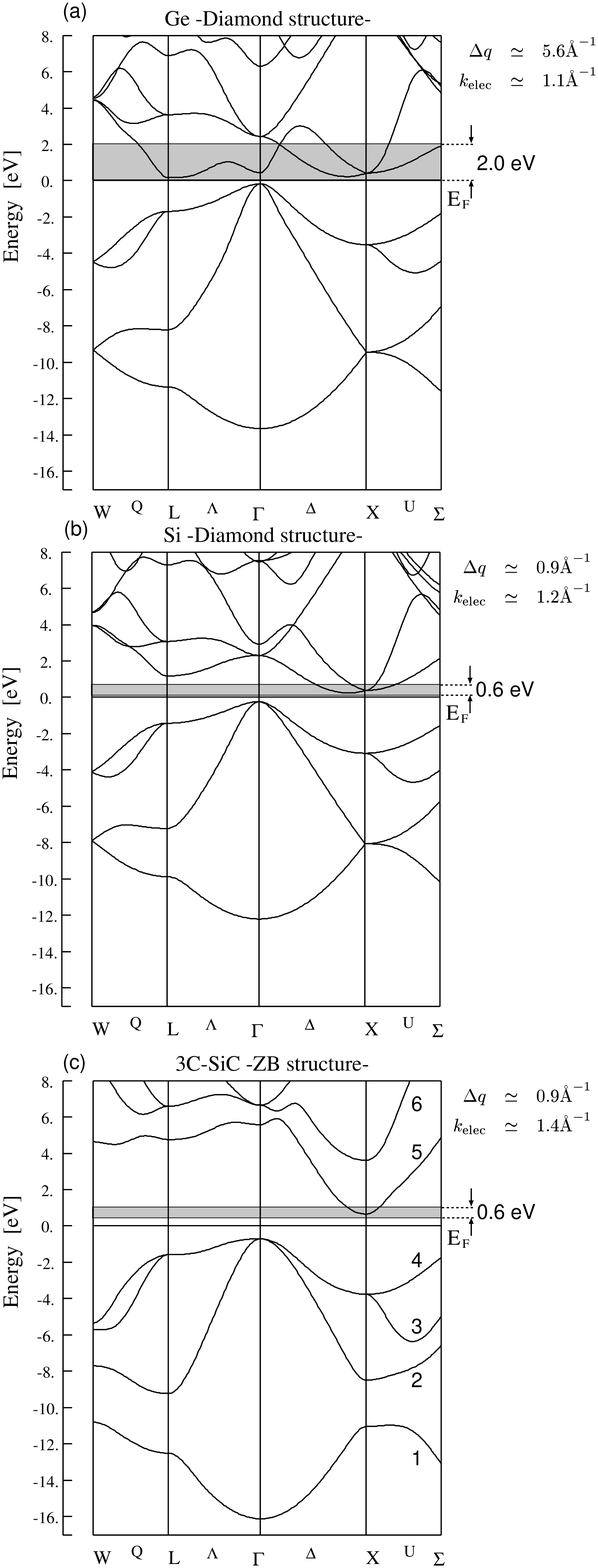}
\caption{\label{fig:band}The band structure of (a)Ge, (b)Si, and
 (c)3C-SiC calculated using density-functional theory (The details
 about the calculation method are described in Sec.\ref{form} B). The shaded
 bars respectively represent the 1$s$ core level widths of Ge (a) and Si (b,c) atoms. The
 numbers attached to the band lines in figure (c) are the band indices.}
\end{figure}

For RIXS at the $K$ edge of Ge we can estimate that $\Delta q\simeq 5.6$ \AA$^{-1}$ and 
$k_{\rm elec}\simeq 1.1$ \AA$^{-1}$
from the $K$ edge energy ($\sim$11100 eV) and the lattice constant.
So, the condition (A) is satisfied.
In Fig.\ref{fig:band}(a) we present the band structure of Ge where the
shaded bar represents the 1$s$ core level width $\Gamma \simeq 2$ eV of Ge.
It is clear that condition (B) is not satisfied.
Therefore, the crystal momentum of the excited electron takes whole　
values in the first Brillouin zone, as does the momentum of the hole
in the final state. Hence the RIXS spectra at the K edge does not show
clear momentum dependence.

The RIXS spectra at the $K$ edge of Si 
has a weak transferred momentum dependence\cite{rf:Maexp},
 which is well reproduced by using our theoretical formulation described in Sec.\ref{form} B. \cite{rf:Sith}
This is because condition (A) and (B) are satisfied for 
RIXS  at the $K$ edge of Si:
(A)$\Delta q\sim k_{\rm elec}$, where $\Delta q\simeq 0.9$ \AA$^{-1}$ and 
$k_{\rm elec}\simeq 1.2$ \AA$^{-1}$ are estimated from 
 the $K$ edge energy ($\simeq$ 1840 eV ) and the lattice constant.
We find (B)$\Gamma < W_{\rm con}$ from Fig. \ref{fig:band}(b)
where the band structure of Si is presented, with the shaded bar representing  
the $1s$ core level width $\Gamma \simeq$ 0.6 eV \cite{rf:Gamma1,
rf:Gamma2}.

Now, we consider the RIXS at the Si $K$ edge of 3C-SiC.
The band structure of 3C-SiC is presented in Fig. \ref{fig:band}(c) with
the shaded bar representing 
the $1s$ core level width $\Gamma \simeq$ 0.6 eV.
Condition (B),$\Gamma \ll W_{\rm con}$, is clearly satisfied.
Condition (A) is also satisfied.
Therefore, we can expect that RIXS at the Si $K$ edge of
3C-SiC has 
strong transferred momentum dependence.
However, to our knowledge, no k-scannable RIXS study near Si $K$ edge 
of 3C-SiC
has been reported, although a pioneering work on the 
band mapping by using k-selective RIXS at the C $K$ edge and Si $L$ edge 
has been performed.\cite{rf:3C-SiC}

It is the purpose of this paper to demonstrate theoretically that k-scannable RIXS can be utilized to study the dispersion of valence bands for broad band materials, taking RIXS at Si $K$ edge of 3C-SiC as a typical example.
We also calculate the RIXS spectra at the C $K$ edge of 3C-SiC and
compare the experimental and theoretical spectra.
This paper is organized as follows.
After describing the formalism for calculation of RIXS spectra in Sec.\
\ref{form}, we investigate transferred momentum dependence of RIXS
spectra in Sec.\ \ref{result} A.
In Sec.\ \ref{result} B and C the incident photon polarization and
energy dependence of spectra are respectively presented.
Sec.\ \ref{result} D is devoted to RIXS near C $K$ edge of 3C-SiC.
Concluding remarks are presented in Section \ref{conc}.
%
%
\section{Formulation}\label{form}
\subsection{Double differential scattering cross-section at $K$ edge}　
RIXS at the $K$ edge is described by a second-order process. The incident photon with energy $\hbar\omega_1$, momentum $\hbar{\bf q}_1$ and polarization ${\bf e}_1$ is virtually　 absorbed by exciting the 1$s$ electron to the conduction band. Then　a photon with energy $\hbar\omega_2$,　momentum $\hbar{\bf q}_2$ and polarization ${\bf e}_2$　 is emitted by filling the core-hole state with a valence electron.
An independent-particle approximation seems to be appropriate,　 since electron correlations are expected to be weak.
Using the generalized Fermi golden rule where the interaction between photons and electrons is treated by second order perturbation theory, we obtain the double differential scattering cross-section as follows;

\begin{eqnarray}\label{dsS}
\frac{d^{2}\sigma}{d\omega_{2} d\Omega_{2}}&\propto&
\sum_{({\bf k},e),({\bf k}^{\prime},h)}\nonumber\\
&\times&\frac
{\left|\sum_{a}\exp(i\Delta{\bf q}\cdot{\bf R}_{a})\overline{t_{a}({\bf k}^{\prime},h|{\bf e}_{2})}
t_{a}({\bf k},e|{\bf e}_{1})\right|^{2}}
{(\epsilon_{e}({\bf k})-\epsilon_{c}-\hbar\omega_{1})^{2}+\Gamma^{2}/4}\nonumber\\
&\times&\delta^{{\bf G}}_{\Delta{\bf q},{\bf k}-{\bf k}^{\prime}}
\delta(\epsilon_{e}({\bf k})-\epsilon_{h}({\bf k}^{\prime})-\hbar\omega),
\end{eqnarray}

where $\hbar\omega=\hbar\omega_{1}-\hbar\omega_{2}$, and
 $\hbar\Delta{\bf q}=\hbar{\bf  q}_{1}-\hbar{\bf q}_{2}$ are the energy and momentum of the
 final state, respectively.
$\epsilon_{e}({\bf k})$, $\epsilon_{h}({\bf k}^{\prime})$ and $\epsilon_{c}$
are the energy of the excited electron with crystal momentum ${\bf k}$
in the conduction band $e$, the energy of the hole with crystal momentum
${\bf k}^{\prime}$ in the valence band $h$, and the energy of the core state
 of a specific atom (Si or C atom in this paper), respectively.
The overline in the equation indicates a complex conjugate.
The $\Gamma$ is the $1s$ core level width of a specific atom.

The quantity 
\begin{equation}\label{DT}
t_{a}({\bf K},b|{\bf e}_{i})\equiv
\int d{\bf r}\overline{\psi_{{\bf K},b}({\bf r})}
{\bf e}_{i}\cdot\hat{{\bf p}}\phi^{1s}({\bf r}-{\bf R}_{a}) 
\end{equation}
describes the transition between the bands and the $1s$ core state 
within the dipole approximation. ${\bf R}_{a}$ is the position vector of a specific atom in the unit cell, 
$\psi_{{\bf K},b}$ is the bloch-wave function of an electron in the band $b$ with crystal momentum ${\bf K}$, and
$\phi^{1s}$ is the 1$s$-atomic orbital of a specific atom.
The crystal momentum conservation for the whole process is contained
in the Kronecker delta,
\begin{equation}\label{MC}
\delta^{\bf G}_{\Delta{\bf q},{\bf k}-{\bf k}^{\prime}}
\equiv  \left\{
\begin{array}{@{\,}ll}
0 & \mbox{: $\Delta{\bf q}-({\bf k}-{\bf k}^{\prime})\notin {\bf G}$ }\\
1 & \mbox{: $\Delta{\bf q}-({\bf k}-{\bf k}^{\prime})\in {\bf G}$ }
\end{array}
\right.,
\end{equation}
where ${\bf G}$ is the set of reciprocal lattice vectors. 
\subsection{Electronic structure calculation of 3C-SiC}
We calculate the electronic structure using the full-potential
linearized augmented-plane-wave (FLAPW) method within the local-density approximation (LDA).
The local exchange-correlation functional of Vosko, Wilk and Nusair is employed.\cite{rf:VWN80} 
The angular momentum in the spherical-wave expansion is truncated at
 $l_{\rm max}=6$ and $7$ for the potential and wave function, respectively.
 The energy cutoff of the plane wave is 12 Ry for the wave function.
Figure \ref{fig:band}(c) shows the calculated band structure of 3C-SiC.  
The numbers attached to bands are band indices used in this paper.
From Fig. \ref{fig:band}(c), we find that 
the energy of band 5 at X point is the minimum energy of the conduction bands.
The obtained band gap $E_{\rm G}^{\rm LDA}$ is 1.3 eV, which is smaller
 than the experimental band gap $E_{\rm G}^{\rm exp}$ $\sim$2.4 eV.
The band gaps are often underestimated in the density-functional calculation with the LDA. 

%
%

\section{Results and discussion}\label{result}
Hereafter, we use the quantity $\hbar\xi_{1}\equiv\hbar\omega_{1}-({\rm
min}_{\bf k}\epsilon_{5}({\bf k})-\epsilon_{c})$, instead of  
the incident photon energy $\hbar\omega_{1}$, to avoid problems with uncertainty in
the value of the core level energy $\epsilon_{c}$, treated as an input parameter in this paper.
$\hbar\xi_{1}$ is the excited electron energy  measured from the bottom
of the conduction band.
In addition, we plot our calculated RIXS spectra 
as functions of $\hbar\omega_1-\hbar\omega_2 - \Delta\epsilon_{\rm G}$
to avoid issues with uncertainty in the value of the experimental band gap, where 
$\Delta\epsilon_{\rm G}=E_{\rm G}^{\rm exp}-E_{\rm G}^{\rm LDA}\simeq$ 1.1 eV.

\subsection{Transferred momentum dependence of RIXS spectra at the Si K edge}
In this section, we show that the RIXS spectra strongly depend on the
transferred momentum by using two kinds of configuration for the scattering.
In one configuration, we can change the magnitude of transferred
momentum whilst keeping its direction fixed and, in the other configuration, we can
change the direction of transferred momentum, whilst keeping its
magnitude fixed.

\begin{figure}[h]
\includegraphics[width=0.85 \linewidth]{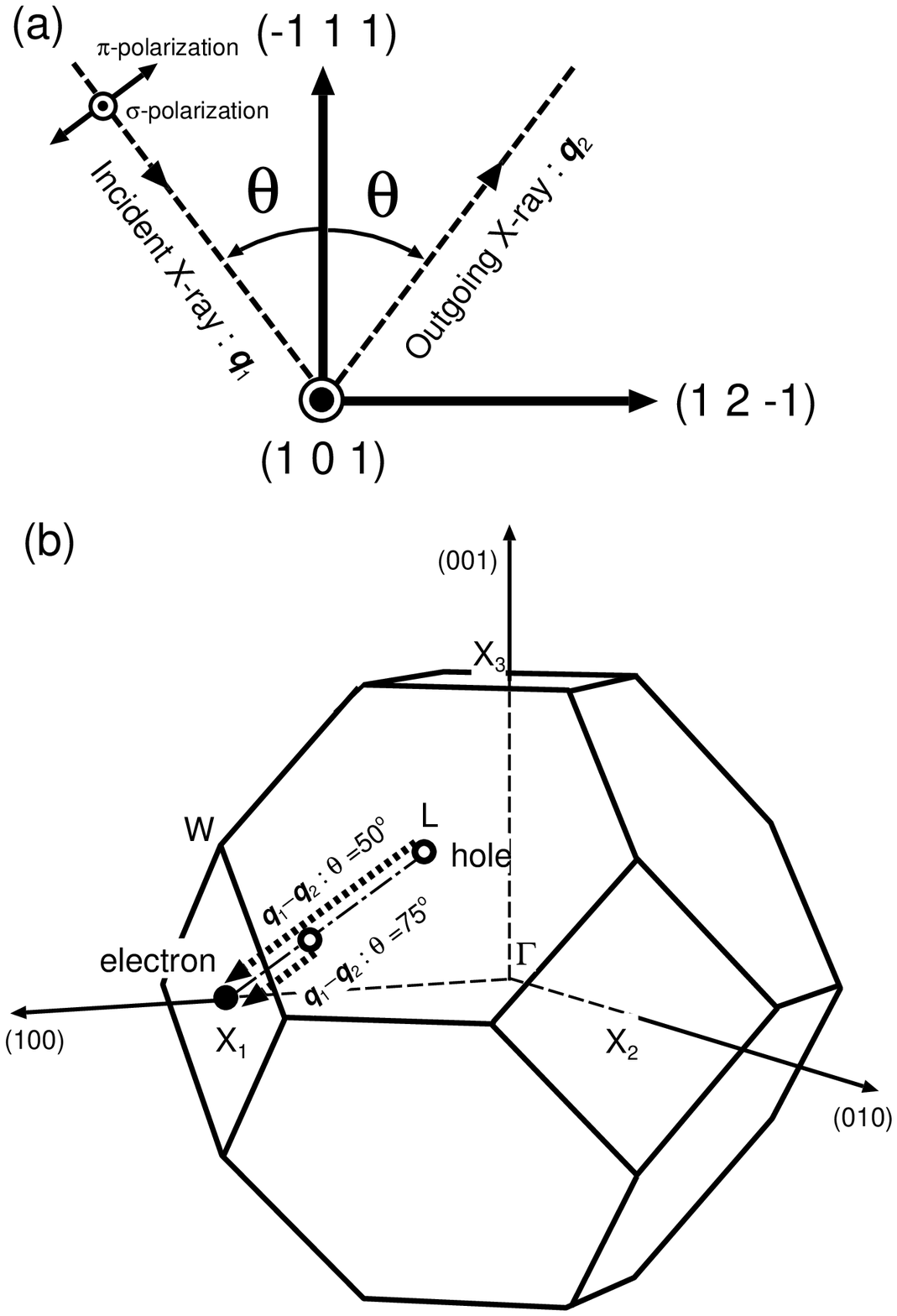}
\caption{\label{fig:setup}(a) The configuration for RIXS. The geometrical relation
 between the crystal axes of 3C-SiC and the momentum of the incident and outgoing
 X-ray are presented. 
 (b) The crystal-momenta of an electron-hole pair in the final state are
 presented in the first Brillouin zone. The dashed arrows represent 
 the transferred momentum for $\theta=50^{\rm \circ}$ and $75^{\rm \circ}$.}
\end{figure}

\begin{figure}
\includegraphics[width=0.8 \linewidth]{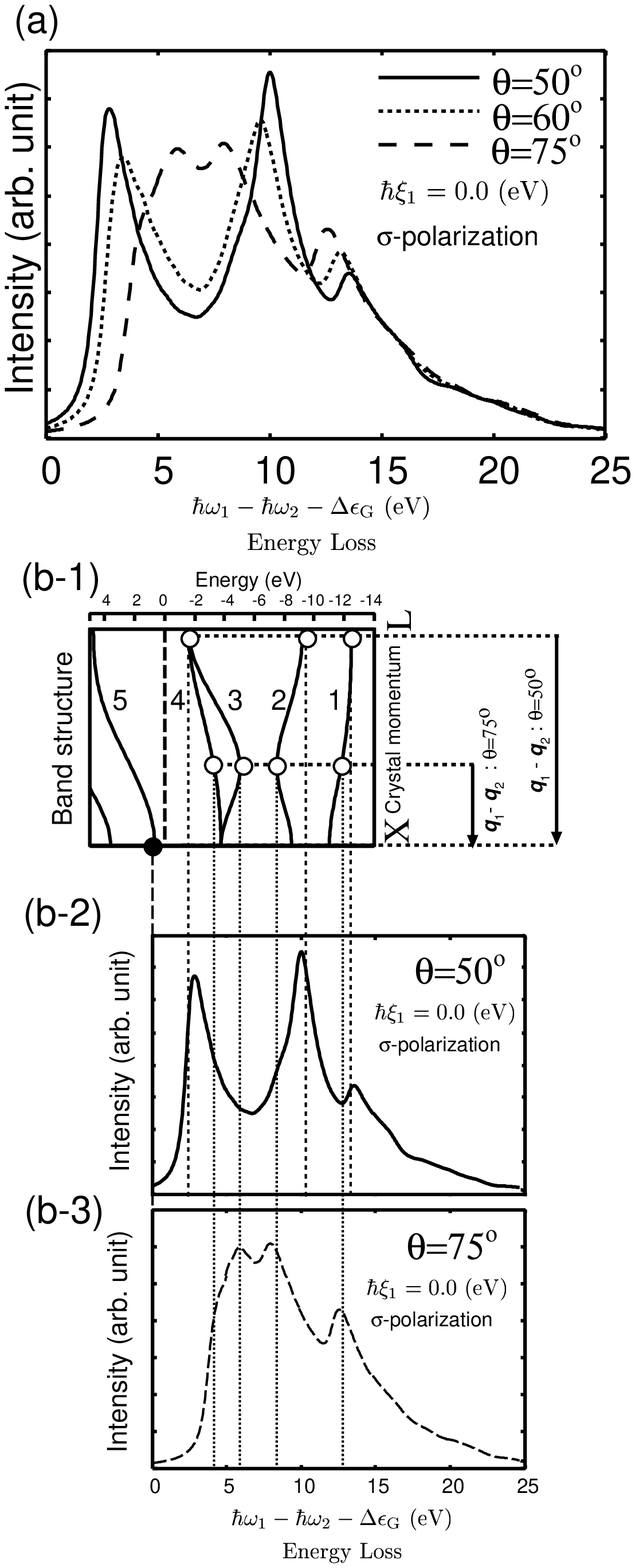}
\caption{\label{fig:k-dep-spec}(a) The transferred momentum dependence of RIXS
 spectra.
 (b) The relation between the transferred momentum dependence of RIXS
 spectra and the valence band dispersion. The vertical dotted and 
dashed lines are guides for the eye: (b-1) The band
 dispersion along the L-X line. The black circle on the conduction band and 
the white circles on the valence bands respectively represent the
excited electron and the hole in the final state for $\theta=50^{\rm \circ}$ and $75^{\rm \circ}$. RIXS spectra for $\theta=50^{\rm \circ}$ (b-2) and $75^{\rm \circ}$(b-3).}
\end{figure}

The first configuration is shown in Fig. \ref{fig:setup}(a).
The magnitude of transferred momentum $\Delta q$ 
can be changed, as $\Delta q \simeq 2|{\bf q}_{1}|\cos\theta$, 
by means of changing the value of $\theta$ without changing its 
direction (1 -1 -1).
The incident photon tuned at the Si K edge 
excites the 1$s$ electron to the X-point ($\hbar\xi_{1}=0.0$ eV). 
The corresponding hole in the final state is created 
according to the law of crystal momentum conservation (Eq.(\ref{MC})).
By changing the value of $\theta$ from $50^{\circ}$ to $75^{\circ}$, 
the crystal momentum of the corresponding hole can be moved 
along the L-X line in the first Brillouin zone, as shown in Fig. \ref{fig:setup}(b), where the black circle represents the excited electron, and 
two dashed arrows and two white circles respectively represent the transferred
momentum and the holes for $\theta=50^{\rm \circ}$ and $75^{\rm \circ}$.
RIXS spectra for $\theta=50^{\circ}$, $60^{\circ}$ and $75^{\circ}$ are shown in Fig.\ref{fig:k-dep-spec}(a), where 
we can clearly see the strong transferred momentum dependence of 
RIXS spectra: two peaks around 6 and 10 eV in the spectrum for $\theta=50^{\circ}$ 
get close to each other and the peak around 14 eV in the spectrum moves
 toward lower energy, when changing $\theta$ from  $50^{\circ}$ to $75^{\circ}$.
From Fig.\ref{fig:k-dep-spec}(b), we can understand 
the relation between the transferred momentum dependence and 
the valence band dispersion as follows.
In Fig.\ref{fig:k-dep-spec}(b-1), the band dispersion along the L-X line
is plotted and the black circle on the conduction band and 
the white circles on the valence bands  respectively represent the
excited electron and the holes in the final state 
for $\theta=50^{\rm \circ}$ and $75^{\rm \circ}$.    
RIXS spectra for $\theta=50^{\rm \circ}$ and $75^{\rm \circ}$ are
presented again in Fig.\ref{fig:k-dep-spec}(b-2) and (b-3), respectively. 
The vertical dashed and dotted lines show that the peaks in the spectra correspond to the energies of the holes.
From Fig.\ref{fig:k-dep-spec}(b-1)-(b-3), we find that 
the peaks around 6, 10 and 14 eV in the spectrum for
$\theta=50^{\rm \circ}$ respectively correspond to 
the holes on the valence bands 4(3),2 and 1.
Therefore, these peaks in the spectrum for
$\theta=50^{\rm \circ}$ move according to the valence band dispersion, which brings with it a transferred momentum dependence of RIXS spectra. 
In the situation mentioned above, 
we expect that the information on the valence band dispersion 
can be extracted from the transferred momentum dependence of the 
RIXS spectra.
%
%

%
%

\begin{figure}
\includegraphics[width=0.8 \linewidth]{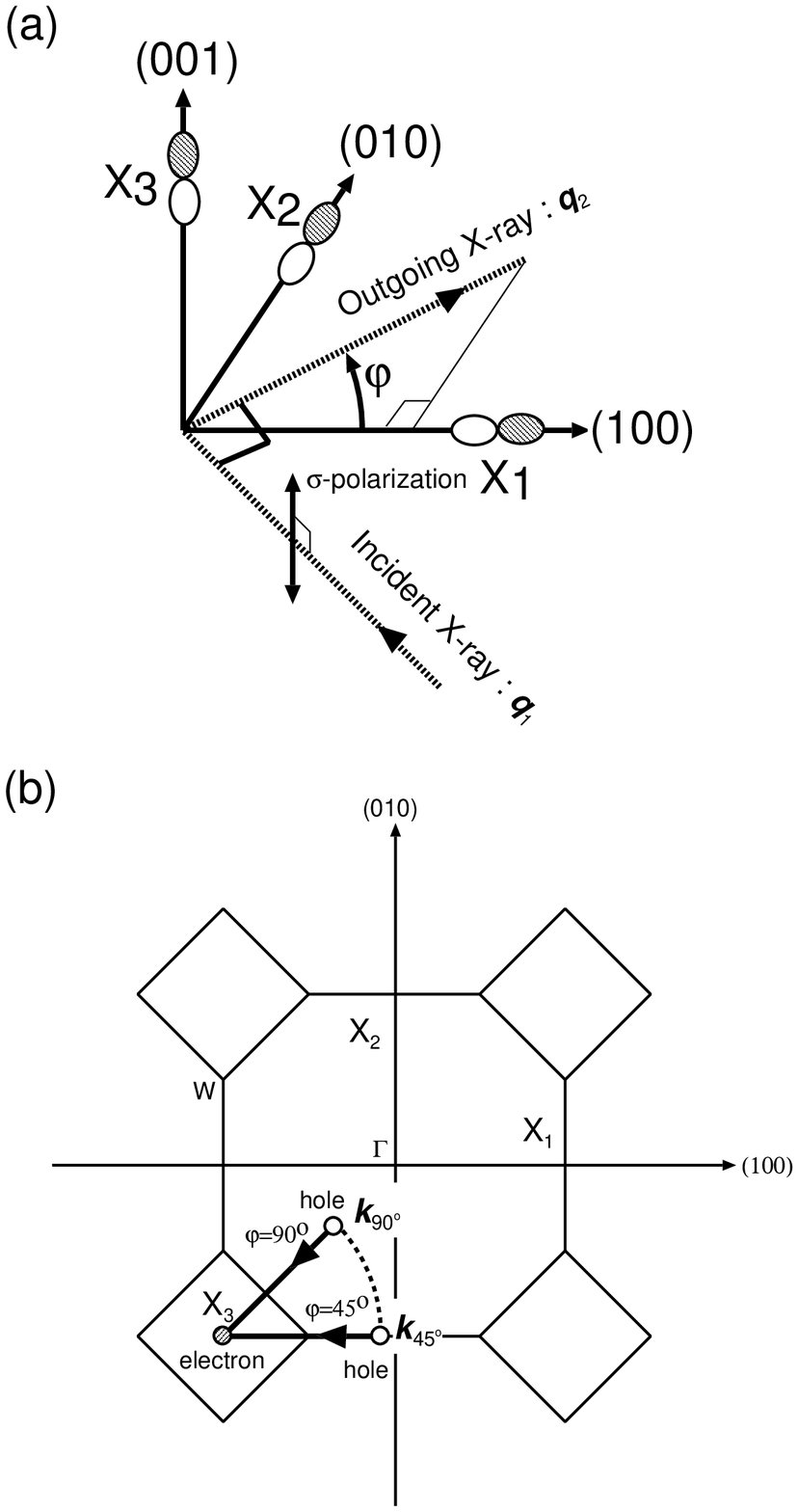}
\caption{\label{fig:aseupt}(a) The configuration for RIXS, where the
 angle between the incident and outgoing momentum is fixed to $90^{\rm
 \circ}$. The geometrical relation
 between the crystal axes of 3C-SiC and the momentum of incident and outgoing
 X-ray are presented. The wave functions of conduction-band 5 at
 X-point, projected onto $p$-symmetric states centering on the Si atom are 
illustrated on the crystal axes. (b)
 The crystal-momenta of the electron-hole pair in the final state are
 presented in the cross-section of the Brillouin zone. The arrows indicate
 the transferred momenta for $\varphi=45^{\rm \circ}$ and $90^{\rm \circ}$.}
\end{figure}

\begin{figure}
\includegraphics[width=0.8 \linewidth]{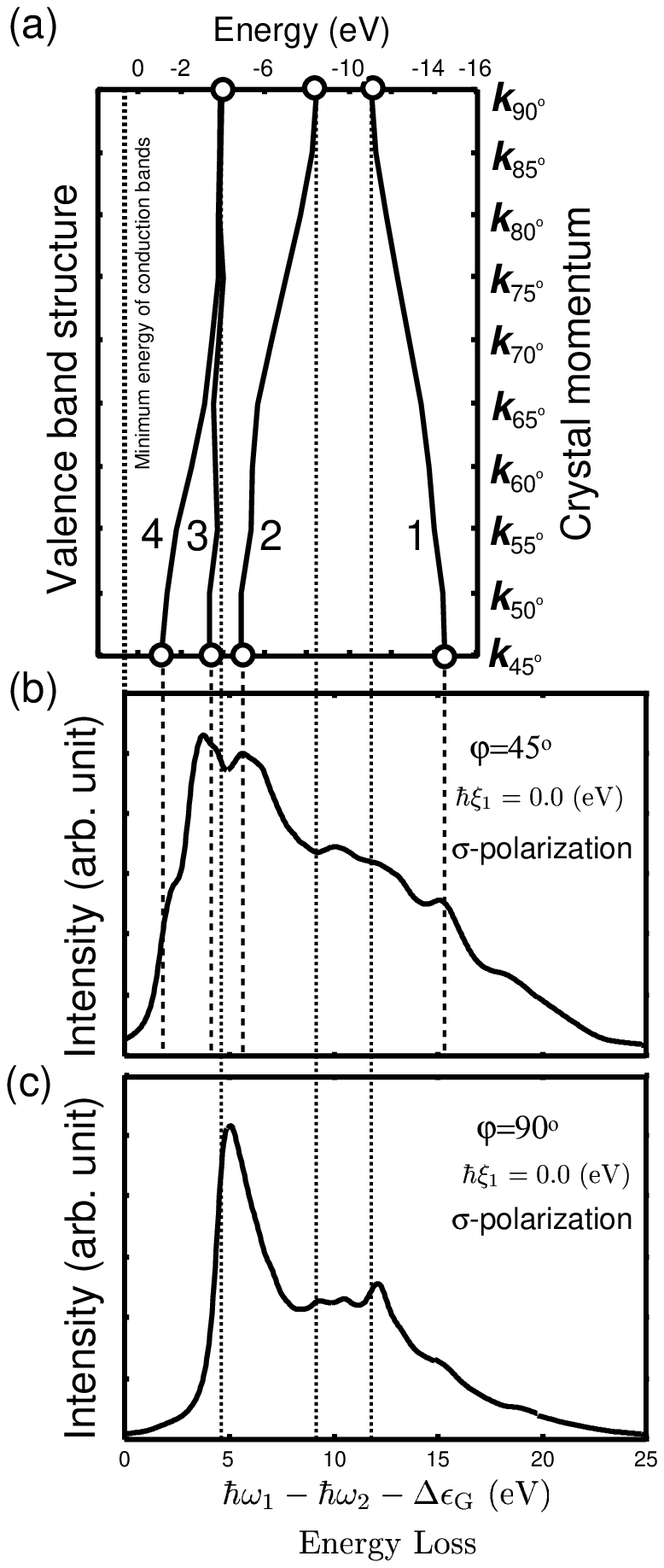}
\caption{\label{fig:k-dep-spec-as} The relation between valence band
 dispersion and spectra. The dotted and dashed lines are guides for the
 eye: (a) The valence band
 dispersion along the dotted arc in Fig.\ref{fig:aseupt}. The white
 circles on the band lines represent the valence band holes for 
$\varphi=45^{\rm \circ}$ and $90^{\rm \circ}$ in the final state. (b). RIXS spectra for $\varphi=45^{\rm \circ}$(b) and $90^{\rm \circ}$(c).}
\end{figure}

We also calculate RIXS spectra using another configuration shown in
Fig. \ref{fig:aseupt}(a).
We can change the direction of the 
transferred momentum by means of rotating the sample around the (001)-axis by $\varphi$,
without changing the magnitude of the transferred momentum 
$\Delta q \simeq \sqrt{2}|{\bf q}_{1}|$.
It seems that this configuration lends itself more easily to experiment than the
configuration shown in Fig. \ref{fig:setup}(a) because we need not change 
the angle between the incoming and the outgoing X-ray.
The incident photon tuned at the Si K edge 
excites the 1$s$-electron to band 5 at the X-point 
where the energy of the conduction band is a minimum
($\hbar\xi_{1}=0.0$ eV).
There are three equivalent X points (X$_1$, X$_2$ and X$_3$ in
Fig.\ref{fig:aseupt}(a)) in the first Brillouin zone.
The wave functions of band 5 at X$_1$, X$_2$ and X$_3$ 
projected onto the $p$-symmetric state centering on the Si atom
are respectively $P_{l=1}\psi_{{\bf k}={\rm X}_{1}, 5}\propto p_{x}$, $P_{l=1}\psi_{{\bf k}={\rm X}_{2}, 5}\propto p_{y}$ and  $P_{l=1}\psi_{{\bf k}={\rm X}_{3}, 5}\propto p_{z}$, as illustrated on the crystal axes in Fig. \ref{fig:aseupt}(a).
(Here, $P_{l=1}$ is an operator of the projection onto the $p$-symmetric
state centering on the Si atom.)
Therefore, the $\sigma$-polarized incident photon tuned at the Si K edge
excites the 1$s$ electron only to the X$_3$-point
according to the selection rule within the dipole approximation ( Eq.(\ref{DT})).
When we change the value of $\varphi$  from $45^{\circ}$ to
$90^{\circ}$, the crystal momentum of the hole corresponding to the
electron at the X$_3$-point 
moves along the dotted
arc in the first Brillouin zone as shown in Fig.\ref{fig:aseupt}(b). 
The black circle in Fig.\ref{fig:aseupt}(b) represents the excited electron, and
two  arrows and two white circles respectively represent the transferred
momentum and the holes for $\varphi=45^{\rm \circ}$ and $90^{\rm \circ}$.
The band dispersion along the dotted arc in Fig.\ref{fig:aseupt}(b)
are presented in Fig.\ref{fig:k-dep-spec-as}(a) where 
the white circles on bands represent the  holes for
$\varphi=45^{\rm \circ}$ and $90^{\rm \circ}$ in the final state. 
The RIXS spectra for $\varphi=45^{\circ}$ and  $90^{\circ}$ are
respectively presented in Fig.\ref{fig:k-dep-spec-as}(b) and (c).
The vertical dashed and doted lines show 
that the peaks in the spectra correspond to the energies of the holes.
From Fig.\ref{fig:k-dep-spec-as}(a-c), we find that the RIXS spectra 
depend on the direction of the  transferred momentum, reflecting 
the valence band dispersion.
%
%

%
%

\subsection{Incident photon polarization dependence of RIXS spectra at the Si K edge}
\begin{figure}[h]
\includegraphics[width=0.85 \linewidth]{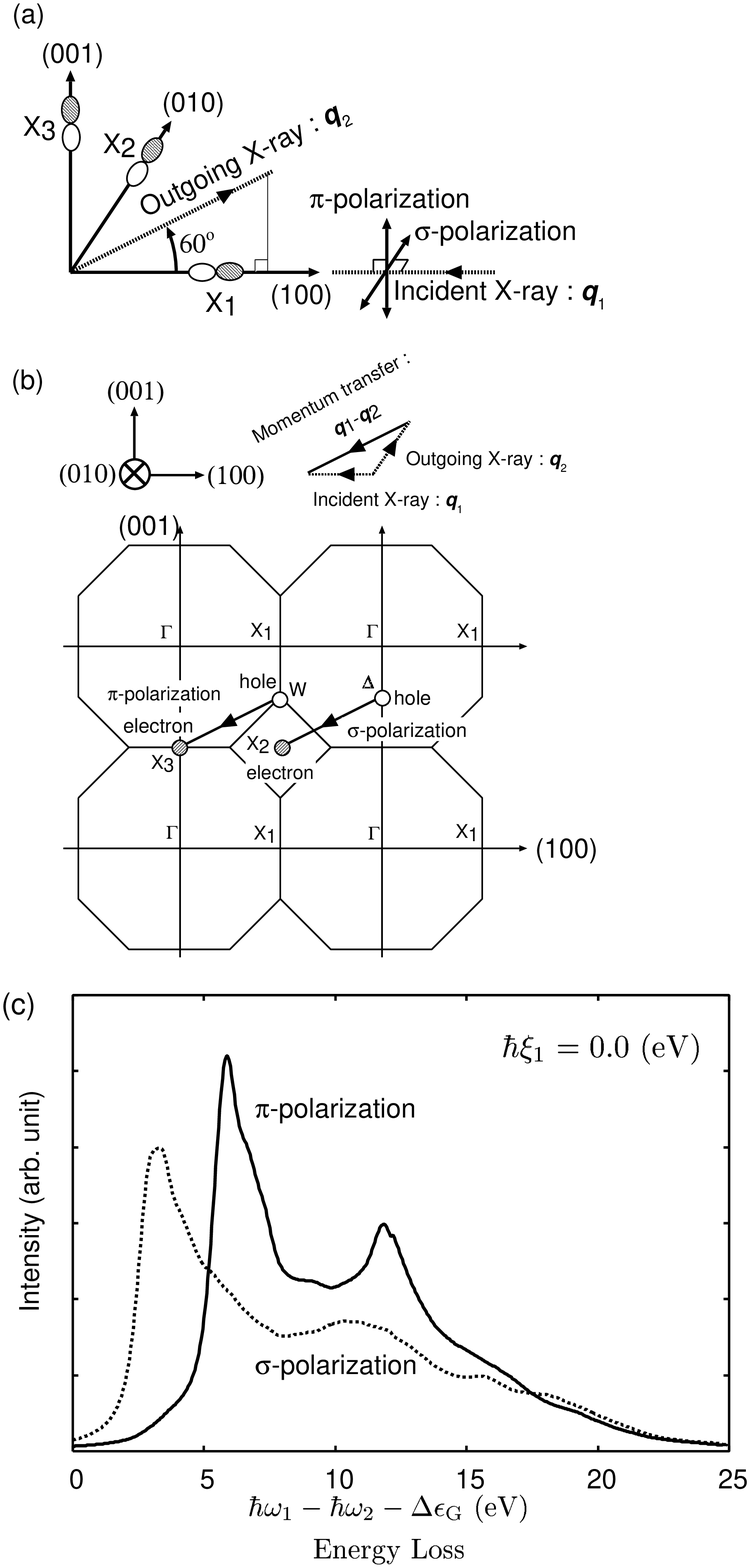}
\caption{\label{fig:pol-dep-set}(a) The configuration for the RIXS to investigate
 the dependence of the incident photon polarization. The geometrical relation
 between crystal axes of 3C-SiC and the momentum of the incident and outgoing
 X-ray are presented. The wave functions of conduction band 5 at
 the X$_1$, X$_2$ and X$_3$-point, projected onto $p$-symmetric states centering on the Si atom are 
illustrated on the crystal axes. (b)For $\pi$ and $\sigma$-polarization, 
the crystal-momenta of the electron-hole pair in the final state are 
presented in the cross-section of k-space. The boundaries of the Brillouin zone
 are also presented. The arrows indicate
 the transferred momenta. (c) The incident photon polarization dependence
 of RIXS spectra.}
\end{figure}

We investigate the incident photon polarization dependence of RIXS
spectra by using a configuration shown in Fig.\ref{fig:pol-dep-set}(a).
On the crystal axes in Fig.\ref{fig:pol-dep-set}(a), we illustrate the
wave functions of band 5 at the X-point projected onto
$p$-symmetric states centering on the Si atom.
From Fig.\ref{fig:pol-dep-set}(a), we find that the incident photon with
$\pi$ ($\sigma$)- polarization tuned at the Si K edge can excite the
$1s$-electron only to the X$_{3}$ (X$_2$) point because of the selection rule within the dipole approximation( Eq.(\ref{DT})).
The corresponding hole in the final state is created at the W ($\Delta$)
point in k-space to satisfy the law of crystal momentum conservation ( Eq.(\ref{MC})), as shown in Fig.\ref{fig:pol-dep-set}(b), where the cross-section of k-space and boundaries of Brillouin zones are also presented.
Therefore, the photon is emitted from the W ($\Delta$) point in the case of 
the $\pi$ ($\sigma$)-polarization and the energy of emitted photons with different polarization is not the same, as shown in Fig.\ref{fig:pol-dep-set} (c). 
The spectra clearly show the strong polarization dependence. For example, the spectrum for $\pi$-polarization has one main peak at approximately 6 eV but the spectrum for $\sigma$-polarization has that at approximately 3 eV.
In the above discussion, it is essential that both $\pi$ and
$\sigma$-polarized incident photons excite the 1$s$ electron to the
equivalent k-points (X$_3$ and X$_2$) but the corresponding holes are
created at the non-equivalent k-points (W and $\Delta$) because of the
sizable momentum transfer in the law of crystal momentum conservation. 
Therefore, the strong polarization dependence is one of the characteristic
features of the k-scannable RIXS which the k-selective RIXS( $\Delta q \ll k_{\rm elec}$ ) does not have. 
We conclude that without changing the transferred momentum, the
electronic structure at non-equivalent k-points can be investigated from
the polarization dependence due to the selection rule and the law of
crystal momentum conservation with a sizable momentum transfer.
%
%

\subsection{Incident photon energy dependence of RIXS spectra near the
  Si K edge}
\begin{figure}[h]
\includegraphics[width=0.8 \linewidth]{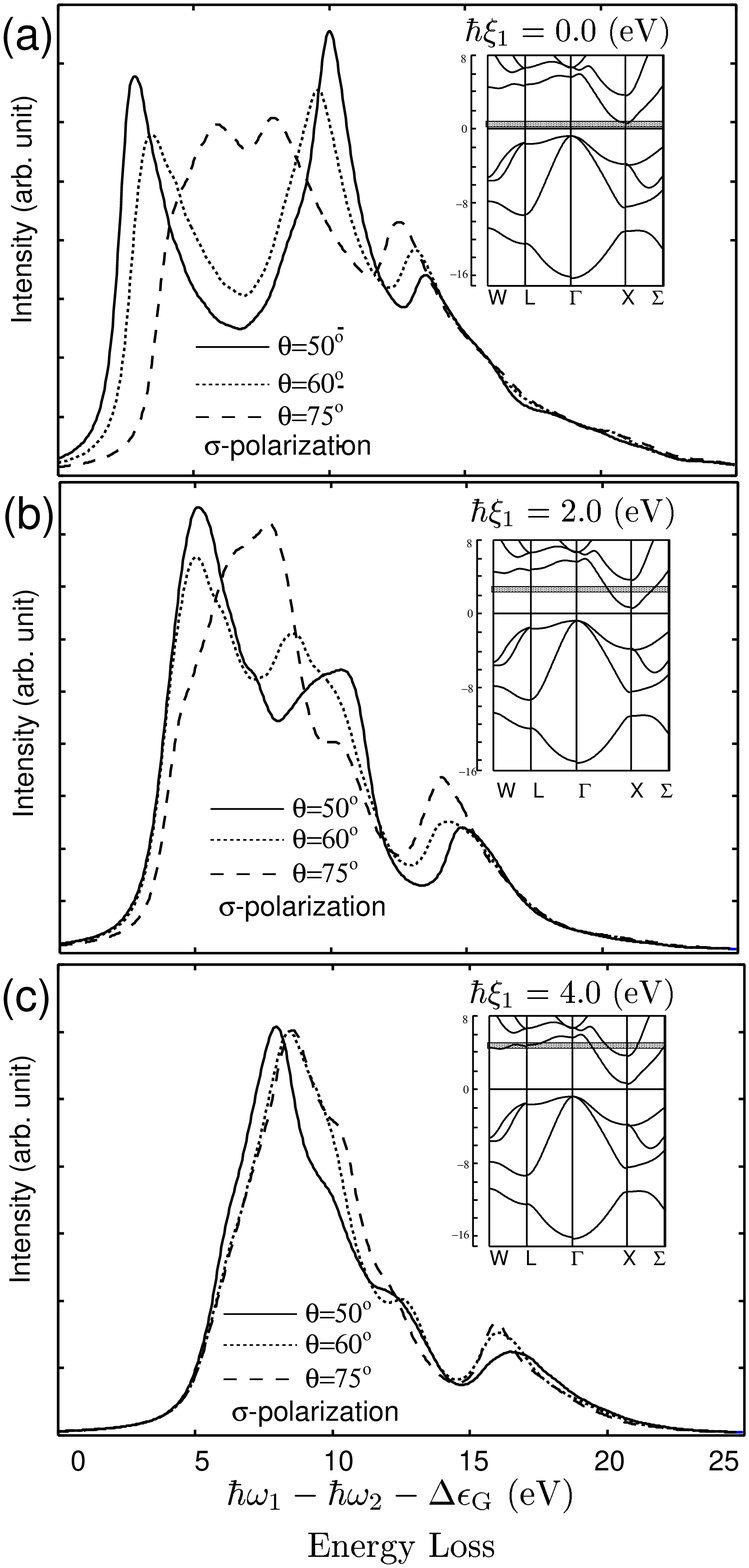}
\caption{\label{fig:ene-dep-spec}The　incident photon energy dependence
 of RIXS spectra near the Si $K$ edge for $\theta=50^{\rm \circ}$,  
$60^{\rm \circ}$ and $75^{\rm \circ}$:
$\hbar\xi_{1}=$ (a)0.0, (b)2.0 and (c)4.0 eV. (Inset) The energy of the excited
 electron is indicated by the shaded bar in the band structure of 
3C-SiC .}
\end{figure}

In Sec.\ \ref{result} A and B, we fixed the incident photon energy only at the
Si K edge ($\hbar\xi_{1}=0.0$ eV).
Here, we investigate the incident photon energy dependence of 
the RIXS spectra by using the configuration shown in Fig. \ref{fig:setup} (a).
The results for $\hbar\xi_{1}$=0.0, 2.0 and 4.0 eV are presented in Fig.\ref{fig:ene-dep-spec}(a)-(c).
The inset in each figure (a)-(c) is the band structure of 3C-SiC and the
shaded bar in the inset indicates the energy of the electron excited by
each incident photon from the 1$s$ core level. 
From Fig.\ref{fig:ene-dep-spec}(a) and (b), the momentum dependence in
the spectra for $\hbar\xi_{1}\simeq$ 0.0 eV is also found in the spectra  for $\hbar\xi_{1}\simeq$ 2.0 eV.
This is because the 1$s$ electron is excited near the X-point and has
selective crystal momenta due to the deep minimum of conduction
band 5 and the relatively narrow 1s core level width, as shown in the inset of Fig.\ref{fig:ene-dep-spec}(b).
Therefore, we may conclude that the features of the spectra are almost
unchanged as far as the excited electron has selective crystal momenta.
On the other hand, if the incident photon energy becomes larger, the crystal momentum of the excited electron can take whole values in the first Brillouin zone as does the momentum of the hole in the final state. 
Therefore, the peak structure of the RIXS spectra is smeared out, and
the spectra have almost no dependence on the transferred momentum.
This is just the case for $\hbar\xi_{1}=4.0$ eV.
The spectra in Fig.\ref{fig:ene-dep-spec}(c) have almost no dependence
on the transferred momentum because the crystal momentum of the excited electron take almost whole values in the first Brillouin zone, as shown in the inset of Fig.\ref{fig:ene-dep-spec}(c).

From the above discussion, we can predict that the spectra have
transferred momentum dependence up to at least $\hbar\xi_{1}=2.0$ eV.

%
%

\subsection{RIXS spectra near the C K edge}
\begin{figure}[h]
\includegraphics[width=0.8 \linewidth]{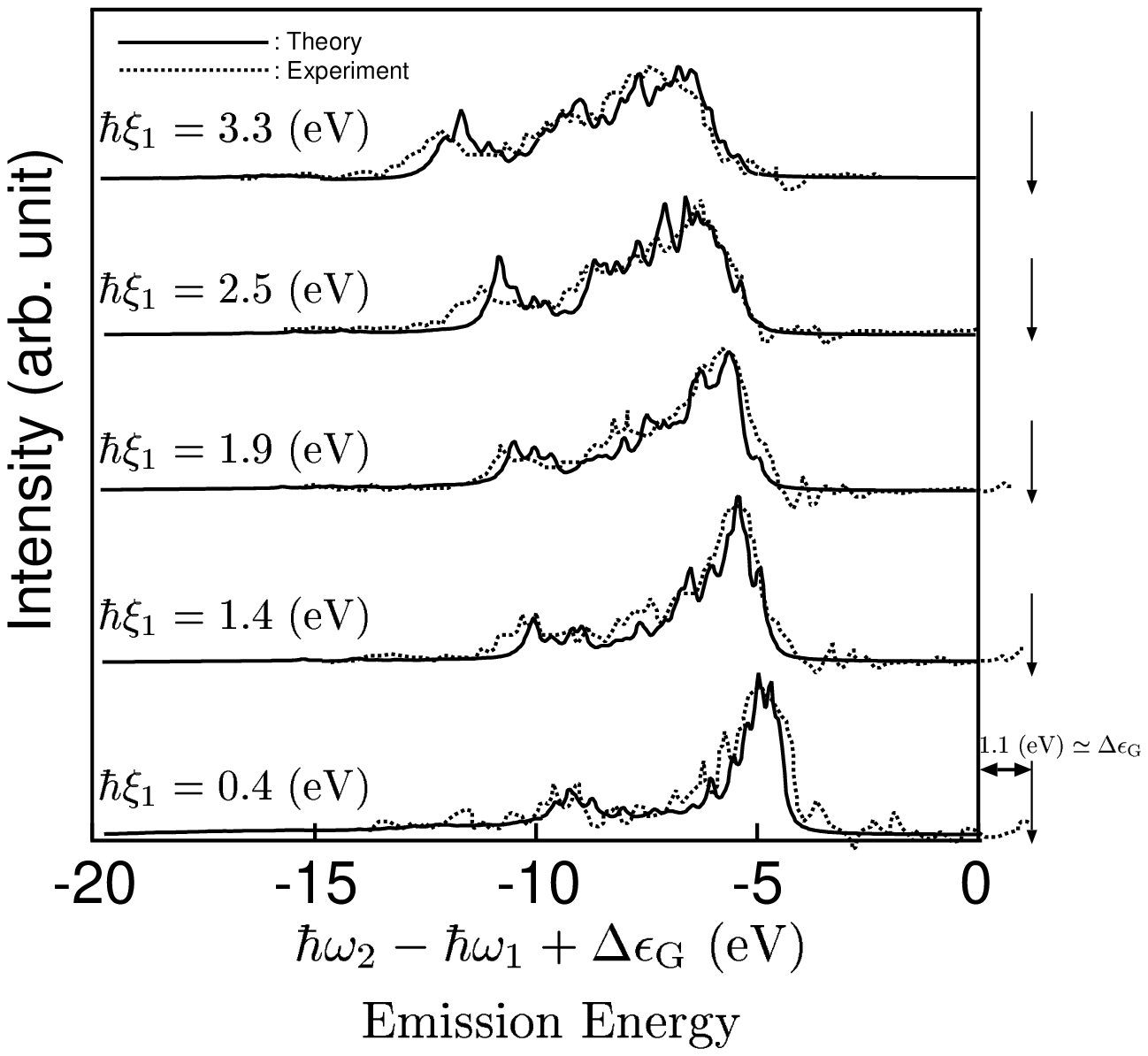}
\caption{\label{fig:C-K-spec}
The comparison between the theoretical (solid
 lines ) and experimental(dotted lines ) RIXS spectra 
near the C $K$ edge. $\hbar\xi_1=$0.4, 1,4, 1.9, 2,5 and 3.3 eV,
 approximately correspond to $\hbar\omega_1=$283.60, 284.60, 285.10, 285.70 and 286.50
 eV in Fig. 1 of the Ref.\cite{rf:3C-SiC}, respectively. Each down-arrow
 indicates the position of the origin of the experimental spectrum plotted as a function of 
$\hbar\omega_{2}-\hbar\omega_{1}$.}
\end{figure}

So far, we have concentrated on a theoretical investigation of the RIXS near the Si $K$ edge.
Last, we calculate the RIXS spectra near the C $K$ edge of
3C-SiC by using the same formulation mentioned in Sec.\ \ref{form}. 
We set the value of the $1s$ core level width of the C atom to 0.1 eV.
The RIXS experiment near the C $K$ edge of 3C-SiC,  
 changing the incident photon energy 
between 283.6 eV and 295.20 eV, has been performed and the experimental RIXS spectra are presented in the Ref.\cite{rf:3C-SiC}.
In Fig.\ref{fig:C-K-spec}, we compare the theoretical and experimental RIXS spectra.
We calculate RIXS spectra as functions of $\hbar\omega_{2}-\hbar\omega_{1}+\Delta\epsilon_{\rm G}$ near the
C $K$ edge for $\hbar\xi_1=$
0.4, 1,4, 1.9, 2,5 and 3.3 eV,
 these energies approximately 
correspond to $\hbar\omega_1=$283.60, 284.60, 285.10, 285.70 and 286.50
 eV in Fig. 1 of the Ref.\cite{rf:3C-SiC}, respectively. 
Each down-arrow in Fig.\ref{fig:C-K-spec} indicates the position of the
origin of the experimental spectrum plotted as a function of $\hbar\omega_{2}-\hbar\omega_{1}$.
As the incident photon energy is increased from $\hbar\xi_{1}=0.4$ eV, the shapes of experimental spectra become broader and the peaks shift toward lower energy.
These features can be well reproduced with our calculations.
However, the position of the lowest-energy peak in each theoretical spectrum is located at slightly higher energy compared with that in each experimental spectrum.
Moreover, the height of the lowest-energy peak in theoretical spectra for 
$\hbar\xi_{1}=$
2.5 and 3.3 eV 
is higher than that in experimental spectra.
This may be because of a linearizing procedure in the FLAPW method,
which becomes less accurate for states highly deviating from the Fermi energy.
Apart from detail, our formulation closely reproduces the experimental RIXS
spectra near the C $K$ edge, which supports our
theoretical prediction of the RIXS at the Si $K$ edge.
Experimental investigation of RIXS at the Si $K$ edge is highly desirable for confirmation of this.
%
%

\section{Concluding remarks}\label{conc}
We have investigated the resonant inelastic X-ray scattering near the
Si $K$ edge of 3C-SiC by systematically changing the
transferred momenta, the incident photon polarization and the incident photon energy, on the basis of an ab initio calculation.
We have demonstrated that the RIXS spectra near the Si $K$ edge have a strong transferred momentum and incident photon polarization dependence.
We have successfully reproduced the experimental RIXS spectra near the
C $K$ edge of 3C-SiC using the same formulation.
%
From the transferred momentum dependence of RIXS spectra, we can extract information on the valence band dispersion.
We can also investigate the valence band at non-equivalent
k-points from the incident photon polarization dependence of the spectra.

We conclude that our results and method are not 
limited to the RIXS at the Si $K$ edge of 3C-SiC.
Similar results are expected in any RIXS
process with a sizable momentum transfer, where the excited 
electron has selective crystal momenta. Therefore, we can conclude 
that a RIXS experiment with a sizable momentum transfer seems
helpful for investigation of valence band dispersion in broad band
materials, considering that RIXS is bulk-sensitive. This is in contrast
to a photo emission with a lower energy photon, which is surface-sensitive.

\begin{acknowledgments}
The authors wish to thank N. Hamada for allowing them to use his FLAPW code,
and A. Agui, M. Mizumaki and J. Igarashi for valuable discussion.
Numerical computation was partly carried 
out using the Computer Facility of Yukawa Institute.
\end{acknowledgments}




\end{document}